\pdfoutput=1
\documentclass[11pt]{article}
\usepackage[preprint]{acl}
\usepackage{times}
\usepackage{latexsym}
\usepackage[T1]{fontenc}
\usepackage[utf8]{inputenc}
\usepackage{microtype}
\usepackage{courier}

\usepackage{graphicx}
\usepackage{amsmath,amsfonts,amssymb}
\usepackage{array}
\usepackage{subcaption}
\usepackage{multirow}
\usepackage{makecell}
\usepackage{booktabs}
\usepackage{refcount}
\usepackage{adjustbox}
\usepackage{float}
\usepackage{tcolorbox}
\tcbuselibrary{breakable}
\hypersetup{
  pdftitle={AcupunctureBench: A Multimodal Benchmark for Evaluating Generative Models in the Acupuncture Domain},
  pdfauthor={Shusheng Li; Yifei Qu; Zichen Han; Xingye Zhang; Zhongxin Wei; Yicheng Qiao; Jiaxin Shen; Changshuai Zhang; Siyuan Guo; Haiyan Ren; Xingfang Pan; Yangyang Liu; Zhaoshui He; Shuguo Hu; Wenjun Tan}
}
\renewcommand{\theHfigure}{main.\arabic{figure}}
\renewcommand{\theHtable}{main.\arabic{table}}
\newcommand{\reusefootnote}[1]{\textsuperscript{\hyperref[#1]{\getrefnumber{#1}}}}

\newtcolorbox{promptbox}{
  breakable,
  fontupper=\small,
  before upper={\raggedright},
  colback=white,
  colframe=black,
  boxrule=0.6pt,
  left=8pt,
  right=8pt,
  top=6pt,
  bottom=6pt,
  before skip=6pt,
  after skip=6pt
}
\newcommand{\prompttitle}[1]{\noindent\textbf{#1}}

\title{AcupunctureBench: A Multimodal Benchmark for Evaluating Generative Models in the Acupuncture Domain}
\author{
\normalsize\bfseries Shusheng Li\textsuperscript{1,2,3}, Yifei Qu\textsuperscript{1,2,3}, Zichen Han\textsuperscript{1,2,3}, Xingye Zhang\textsuperscript{1,2,3} \\
\normalsize\bfseries Zhongxin Wei\textsuperscript{1,2,3}, Yicheng Qiao\textsuperscript{1,2,3}, Jiaxin Shen\textsuperscript{1,2,3}, Changshuai Zhang\textsuperscript{1,2,3} \\
\normalsize\bfseries Siyuan Guo\textsuperscript{1,2,3}, Haiyan Ren\textsuperscript{4}, Xingfang Pan\textsuperscript{4}, Yangyang Liu\textsuperscript{4} \\
\normalsize\bfseries Zhaoshui He\textsuperscript{5}, Shuguo Hu\textsuperscript{6}, Wenjun Tan\textsuperscript{1,2,3}\thanks{\raggedright Corresponding author: \texttt{tanwenjun@cse.neu.edu.cn}.} \\[0.4em]
\small\normalfont\textsuperscript{1}Key Laboratory of Intelligent Computing of Medical Images, Ministry of Education,\\
\small\normalfont Northeastern University, China.\\
\small\normalfont\textsuperscript{2}School of Computer Science and Engineering, Northeastern University, Shenyang 110819, China\\
\small\normalfont\textsuperscript{3}National Frontiers Science Center for Industrial Intelligence and Systems Optimization,\\
\small\normalfont Northeastern University, China\\
\small\normalfont\textsuperscript{4}Research Center of Experimental Acupuncture Science,\\
\small\normalfont Tianjin University of Traditional Chinese Medicine, Tianjin 301617, China\\
\small\normalfont\textsuperscript{5}School of Automation, Guangdong University of Technology, Guangzhou 510006, China\\
\small\normalfont\textsuperscript{6}School of Computer Science and Technology, Harbin Institute of Technology (Shenzhen),\\
\small\normalfont Shenzhen 518055, China}

\begin{document}
\maketitle


\begin{abstract}
As large language models (LLMs) are increasingly applied in the field of traditional Chinese medicine (TCM), standardized benchmark evaluations for TCM have gradually emerged. Acupuncture is a crucial component of TCM. However, existing evaluation benchmarks are lacking in this domain. To address this gap, this paper proposes AcupunctureBench, the first independently constructed large-scale multimodal benchmark for the acupuncture domain, covering text, image, and video modalities. The AcupunctureBench spans multidimensional knowledge in acupuncture, including acupuncture theory, meridians and acupoints, clinical reasoning, TCM syndromes, and practical acupoint operation. To achieve large-scale construction while ensuring medical professionalism, this work designs a human-machine collaborative data construction pipeline and proposes efficient deduplication strategies to reduce redundant samples, while extensively adopting data synthesis to mitigate data leakage risks. AcupunctureBench contains a total of 58,526 question samples and 1,216 video samples, covering multiple task types including objective questions, subjective question answering, case analysis, image understanding, image reasoning, and video understanding. Based on this benchmark, this paper evaluates 14 large language models, 14 large multimodal models, and 4 video foundation models, and further releases the first dedicated evaluation website for the acupuncture domain.
\end{abstract}

\begin{figure*}[!t]
\centering
\includegraphics[width=\textwidth]{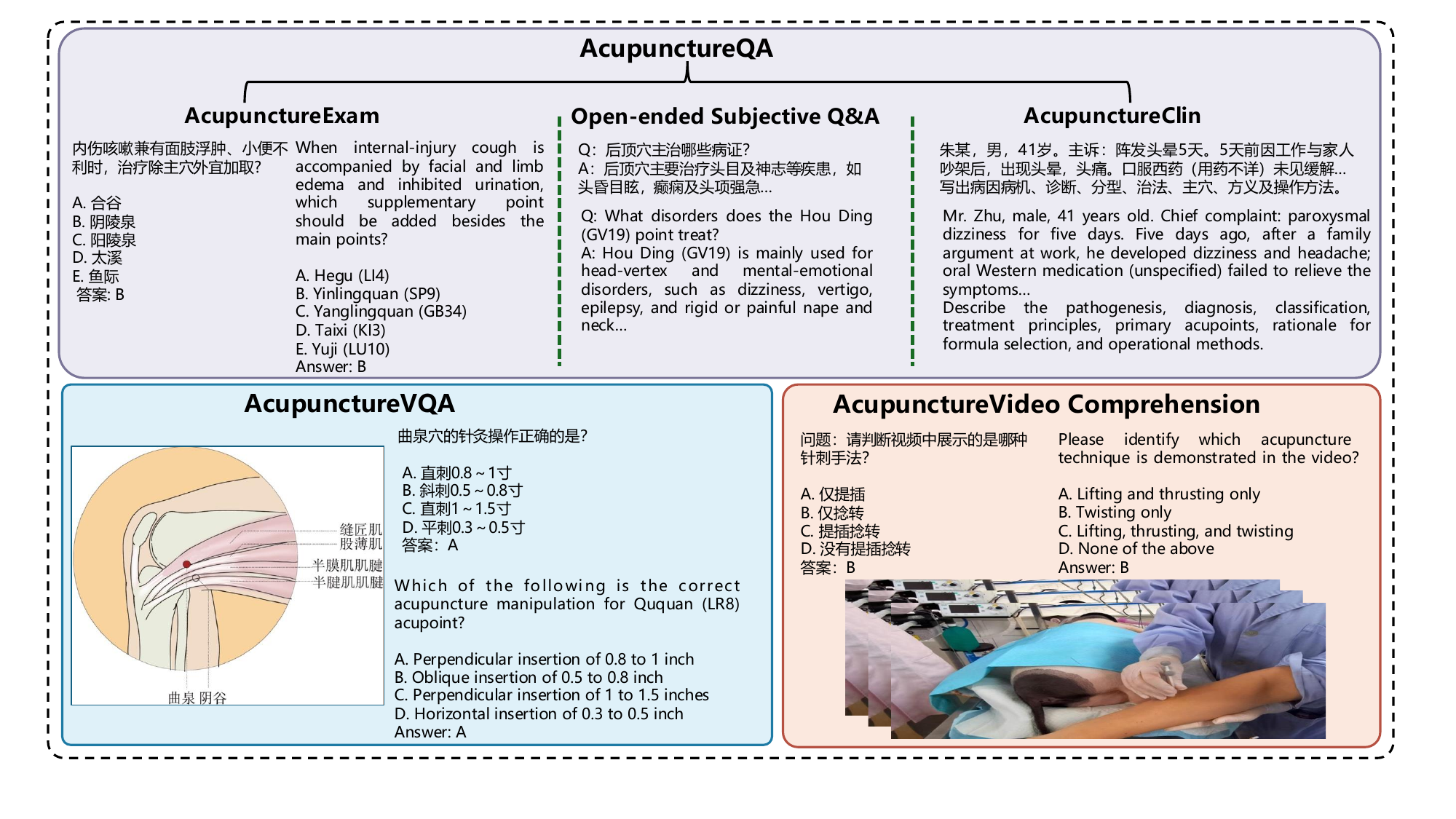} 
\caption{ AcupunctureBench consists of three data subsets: AcupunctureQA, AcupunctureVQA, and AcupunctureVideo. Each subset provides representative data examples.
}
\label{fig:overview}
\end{figure*}

\section{Introduction}
\label{sec:introduction}
Acupuncture is a traditional medical intervention grounded in meridian theory, which regulates physiological functions through needle insertion or moxibustion at specific acupoints. In recent years, its scientific validity has been increasingly supported by modern experimental studies. For example, a study published in \textit{Nature} provided reproducible neuroimmunological evidence for the biological mechanisms of acupuncture based on animal models\cite{liu2021neuroanatomical}. Clinically, its therapeutic efficacy within modern healthcare systems has been demonstrated\cite{vickers2012acupuncture, vickers2018acupuncture}. As an integral component of Traditional Chinese Medicine (TCM), acupuncture is deeply interwoven with TCM theories concerning meridians, internal organs, and pattern differentiation. However, its practice is highly dependent on anatomical localization, visual assessment, and temporal techniques, distinguishing it from the prescription of herbal medicine. A systematic evaluation of large  models in the context of acupuncture can help delineate the boundaries of their capabilities in professional knowledge comprehension and multimodal reasoning, thereby providing a scientific foundation for their safe application in medical education and clinical decision support.

Existing text-based benchmark datasets in the field of TCM can be broadly categorized into three groups. The first category comprises comprehensive TCM capability evaluation benchmarks, which are designed to assess the extent to which LLMs cover and understand TCM knowledge. Representative benchmarks include TCMBench\cite{yue2024tcmbench}, TCM-3CEval\cite{huang2026triaxial}, TCM-Eval\cite{cheng2025tcm}, and MTCMB\cite{kong2025mtcmb}. The second category focuses on syndrome differentiation benchmarks, which evaluate LLMs’ ability to comprehend and apply TCM diagnostic reasoning. Typical examples include ATCMD\cite{lin2026atcmd}, TCMEval-SDT\cite{wang2025tcmeval}, and TCM-SD\cite{mucheng2022tcm}. The third category consists of knowledge retrieval–oriented benchmarks, such as TCMD\cite{yu2024tcmd}, which are primarily used to assess model performance in TCM knowledge-based question answering and factual retrieval tasks. Overall, existing TCM text benchmarks have established a relatively comprehensive evaluation framework encompassing theoretical knowledge, syndrome reasoning, and integrated assessment. However, they share three common limitations: (1) a heavy reliance on textual modalities, with insufficient modeling of other modalities; (2) limited attention to the domain of acupuncture and moxibustion; and (3) a lack of evaluation strategies that reflect real world clinical practice. These limitations directly constrain the effectiveness of current benchmarks in evaluating LLMs’ capabilities in the acupuncture domain. To address these issues, TCM-Ladder\cite{xie2025tcm} introduces the first multimodal evaluation benchmark for the TCM field, however, it remains largely focused on the domain of herbal medicine. Although several studies\cite{li2025taming,liu2025evaluating} have explored evaluation methods for acupuncture-related tasks, no dedicated benchmark dataset has yet been proposed for systematically evaluating LLMs in the acupuncture domain.

To address the challenges of data scarcity and limited modality coverage in evaluating large models within the acupuncture domain, we propose AcupunctureBench, as illustrated in Figure~\ref{fig:overview}, the first large-scale multimodal benchmark specifically designed for foundation models focused on acupuncture. AcupunctureBench is entirely self-constructed and does not rely on any post-processing or reorganization of existing public datasets. It covers three modalities: text (AcupunctureQA), visual question answering (AcupunctureVQA), and video (AcupunctureVideo).

AcupunctureQA consists of three types of tasks: objective multiple-choice questions, subjective open-ended questions, and clinical case analysis. The objective questions and case analysis tasks are derived from authoritative textbooks rather than fragmented internet sources, ensuring high data quality. Specifically, the objective questions include 1,735 samples spanning three major categories: meridians and acupoints, acupuncture and moxibustion techniques, and TCM syndromes. The case analysis subset contains 100 samples, each systematically annotated across five dimensions, including etiology and pathogenesis. For subjective question–answer pairs, although expert physicians can produce high-quality annotations, their limited number and domain-specific focus make it difficult to ensure both diversity and scale. To overcome this limitation, we leverage high-performing LLMs to generate initial QA pairs, which are then carefully reviewed and refined by multiple professional physicians. This collaborative human–LLM pipeline effectively balances data quality and quantity. In addition, we introduce a hash table-based deduplication strategy to efficiently remove redundant QA pairs. The resulting subjective QA set comprises 45,962 samples across four primary domains: acupuncture, TCM syndromes, moxibustion, and tuina.

When constructing the acupuncture-specific VQA benchmark, we encountered significant noise and data scarcity in internet-sourced acupuncture images. To address this issue, we use textbook illustrations as the primary data source and employ both licensed acupuncturists and LLMs for annotation. The resulting AcupunctureVQA dataset contains 10,729 samples, covering two task types: image understanding and image reasoning. We further construct AcupunctureVideo, a video benchmark focused on practical acupuncture procedures, consisting of 1,216 samples divided into three categories: (1) standardized demonstrations of human acupoint needling performed by professional acupuncturists, (2) real-world clinical procedure videos recorded in authentic treatment settings, and (3) standardized acupuncture demonstrations on experimental pigs conducted by trained practitioners. Finally, we conduct a comprehensive evaluation on AcupunctureBench, benchmarking 14 large language models (LLMs), 14 large multimodal models (LMMs), and 4 video foundation models (ViFMs), providing a systematic assessment of multimodal model capabilities in the acupuncture domain.

\section{Methods}
\label{sec:methods}
\subsection{AcupunctureQA}
The AcupunctureQA dataset consists of three components: AcupunctureExam, subjective QA pairs, and AcupunctureClin.  AcupunctureExam comprises the objective question section with 1,735 sample questions covering three major categories: meridians and acupoints, acupuncture and moxibustion techniques, and TCM syndromes.  AcupunctureClin includes 100 sample case analyses, each annotated across five dimensions, including pathogenesis and disease mechanisms.  The above detailed descriptions are provided in the Appendix.  In addition, the subjective question–answer component contains 45,962 QA pairs.  The detailed description of the subjective QA questions is presented below.

\subsubsection{Data Processing}
Two heterogeneous LLMs ERNIE-X1\cite{sun2019ernie} and DeepSeek-R1\cite{guo2025deepseek} were employed to generate QA pairs in accordance with the designed prompt, the prompt is shown in \hyperref[app:qa-generation]{Subsection I} of Section~\ref{app:prompts}. Subsequently, to efficiently eliminate redundant QA pairs from above dataset, we introduced a hash table-based deduplication strategy denoted as $\mathcal{H}$, aimed at improving throughput in large-scale data processing scenarios. The original QA set $\mathcal{Q}$ was duplicated into a candidate set $D_{\text{cand}}$ and a reference set $D_{\text{ref}}$ to facilitate intra-dataset duplicate detection.

In comparison, traditional deduplication methods typically rely on pairwise comparisons:

\[
T_{\text{naive}} = \sum_{i=1}^{N} \sum_{j=1}^{M} \text{F}_c(q^{(i)}, q^{(j)}),
\]

where $\text{F}_c(\cdot,\cdot)$ denotes the exact match operation between two QA pairs. This approach incurs a time complexity of $\mathcal{O}(N \cdot M)$, which is computationally expensive at scale. To reduce this computational overhead, we computed a hash value for each QA pair $q^{(j)} \in D_{\text{ref}}$, denoted as $k_j = h(q^{(j)})$, and constructed a hash lookup table $
\mathcal{T}_{\text{ref}} = \{ k_j \mid q^{(j)} \in D_{\text{ref}} \}$, requiring a linear traversal over $M = |D_{\text{ref}}|$ samples with a time complexity of $\mathcal{O}(M)$. Next, for each QA pair $q^{(i)} \in D_{\text{cand}}$, we computed its hash key $k_i = h(q^{(i)})$, and determined duplication status using the indicator function $\delta_i = \mathbb{I}[k_i \in \mathcal{T}_{\text{ref}}]$, where $\mathbb{I}[\cdot]$ is the indicator function. Assuming uniform hashing, the average lookup time in the hash table is $\mathcal{O}(1)$, yielding a total deduplication complexity of $\mathcal{O}(N)$ for $N = |D_{\text{cand}}|$ candidate samples. Therefore, the overall time complexity of the proposed strategy is $ T_{\mathcal{H}} = \mathcal{O}(M + N)$, which significantly reduces computational cost compared to the traditional $\mathcal{O}(N \cdot M)$ method. By filtering out all duplicate QA pairs satisfying $\delta_i = 1$, we obtained the final deduplicated QA dataset, comprising approximately 53,971 QA samples:

\[
D_e = \{ q^{(i)} \in D_{\text{cand}} \mid \delta_i = 0 \}.
\]

Subsequently, high-quality filtering was performed. First, the QWQ-32B-Preview\footnote{\url{https://qwenlm.github.io/blog/qwq-32b-preview/}} was employed to score each QA pair based on the rubric defined in Table~\ref{tab:assessment-rubric} of the Appendix. A total of 49,157 QA pairs with scores above 8 (out of 10) were retained. In the second stage, 10 experts with master’s degrees or above in acupuncture and tuina independently evaluated all QA pairs using the same rubric. In the final stage, 10\% of the QA pairs were randomly sampled and further validated by two professors in acupuncture and tuina to ensure data quality. Ultimately, a curated set of 45,962 high quality QA pairs was obtained.

\begin{figure*}[!t]
    \centering
    \captionsetup[subfigure]{labelformat=parens} 
    \captionsetup[subfigure]{labelfont={sf}}  
 
    \begin{subfigure}[t]{0.49\textwidth}
        \centering
        \includegraphics[width=\textwidth]{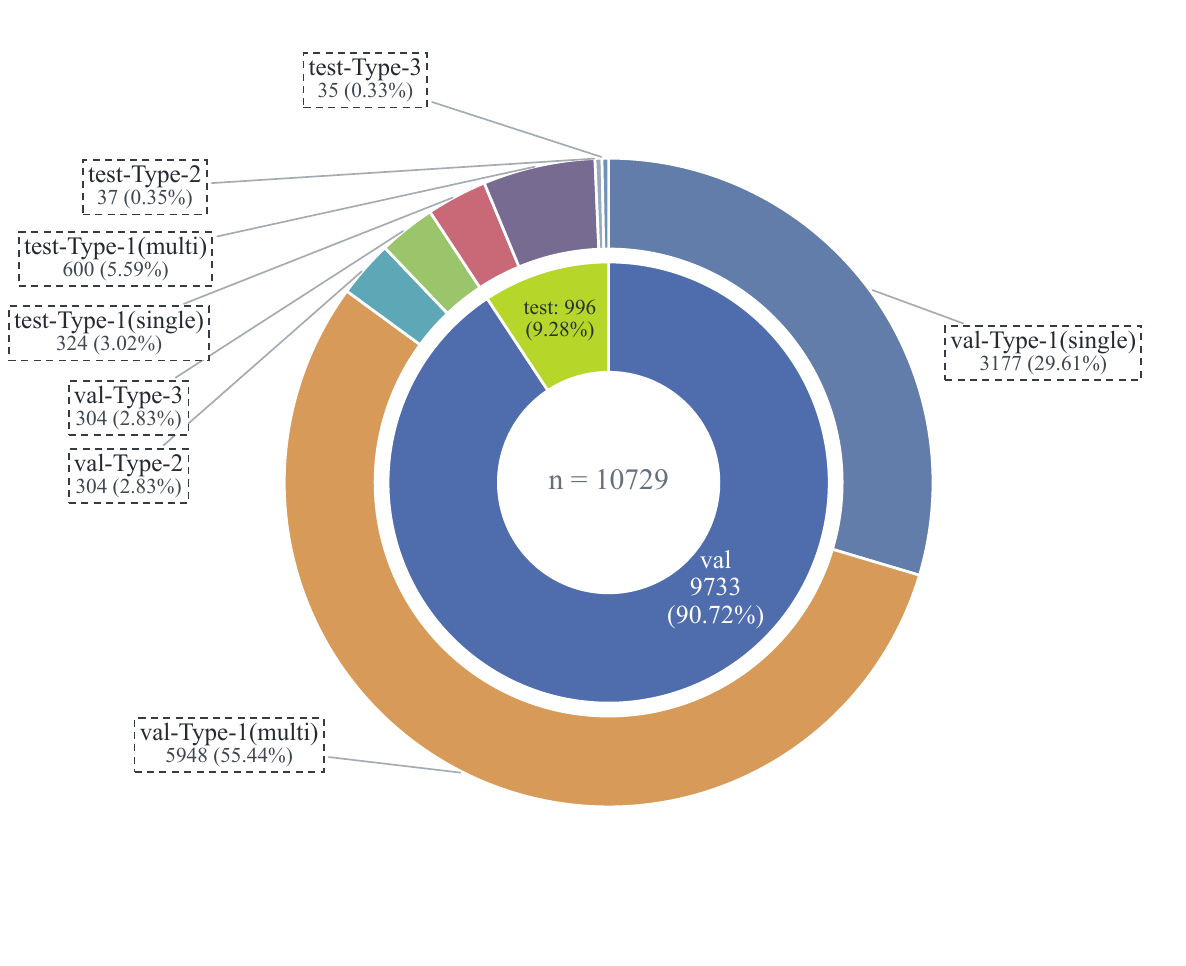}
\phantomsubcaption
        \label{fig:vqa-first}
    \end{subfigure}
    \hfill
    \begin{subfigure}[t]{0.49\textwidth}
        \centering
        \includegraphics[width=\textwidth]{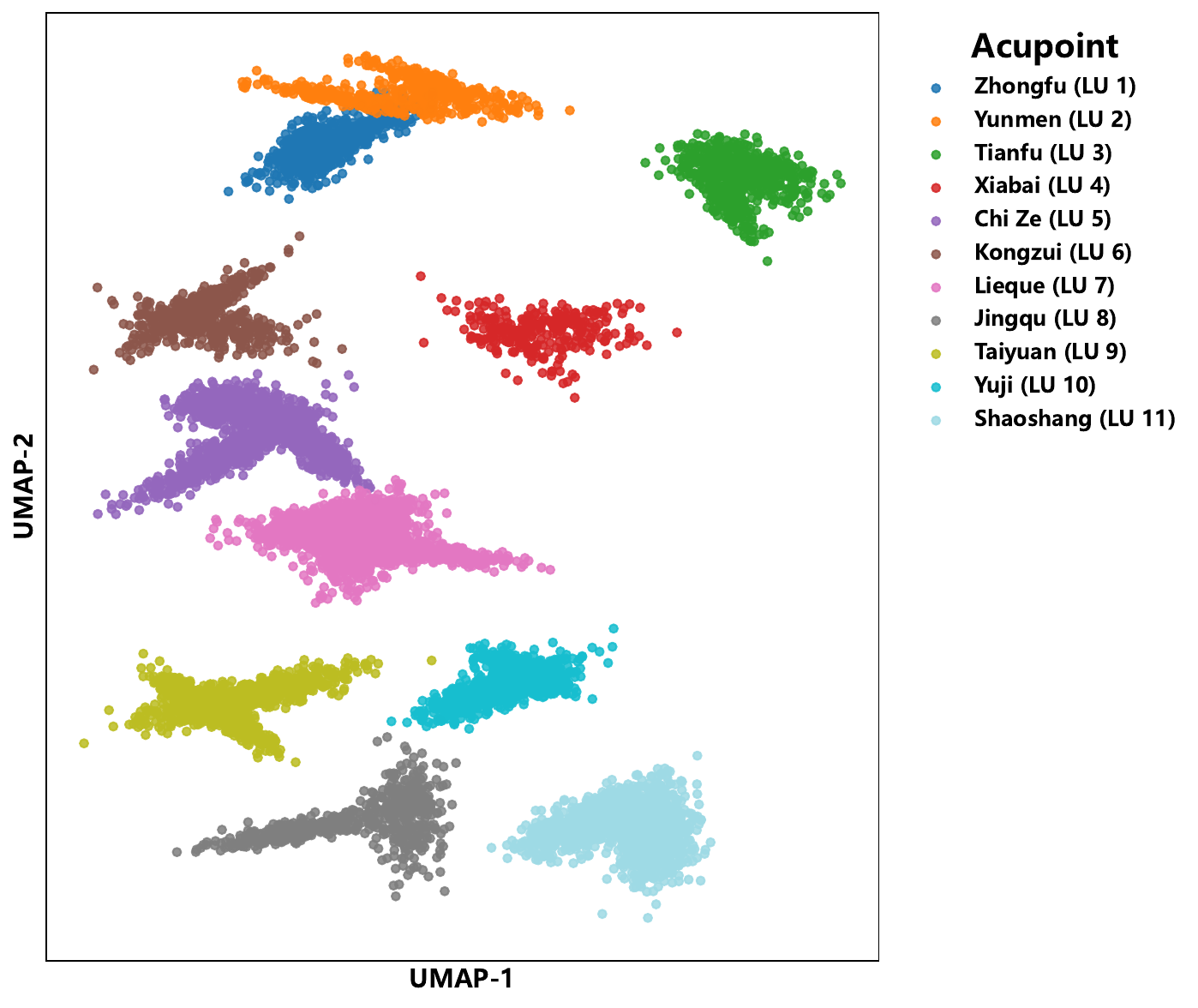}
\phantomsubcaption
        \label{fig:vqa-second}
    \end{subfigure}
    
    \caption{The left panel displays the visualization of AcupunctureVQA data statistics, whereas the right panel presents the statistical visualization of each item corresponding to all acupoints under the Hand Taiyin Lung Meridian category in the AcupunctureVQA dataset.} 
    \label{fig:vqa-statistics}
\end{figure*}

\subsubsection{Data Statistics}
A total of 45,962 diverse QA pairs were constructed across four primary domains: acupuncture, TCM syndromes, moxibustion, and tuina, as summarized in Figure~\ref{fig:app-qa-statistics} of the Appendix. Initially, we generated 17,781 QA pairs related to the field of acupuncture. To evaluate the diversity of this dataset, we designed and applied an acupoint lexicon to perform keyword matching within the QA content. The results showed that the dataset covered the vast majority of entries in the acupoint lexicon, involving approximately 350 distinct acupoints, thereby demonstrating substantial content diversity.

Given that acupuncture is predominantly used in clinical therapeutic contexts, we further developed syndrome-oriented professional dialogue data. To assess the diversity and coverage of this dialogue dataset, a dictionary of TCM syndromes was constructed and employed for matching analysis. The results indicated that the disease-focused QA set encompassed most entries in the TCM disease lexicon, covering approximately 380 disease types, and totaling 26,121 disease-related QA pairs. Each data instance in the otorhinolaryngology and sensory organ category within the TCM syndromes domain was visualized, as shown in Figure~\ref{fig:app-qa-statistics} of the Appendix.

Finally, recognizing that moxibustion and tuina are important therapeutic modalities closely associated with acupuncture, we integrated them into the overall data construction framework. Accordingly, 644 QA pairs were generated for the moxibustion domain and 1,416 for the tuina domain.

\subsection{AcupunctureVQA}

\subsubsection{Dataset Construction}

Two task types, Image Understanding and Image Reasoning, were designed to evaluate the performance of LLMs in the domain of acupuncture. Specifically, the Image Understanding task assesses the model’s ability to identify which acupoints are present in a given image and to locate specified acupoints. The Image Reasoning task evaluates the model’s reasoning capability regarding how a specified acupoint is manipulated within the image.

For the acupoint localization questions in the Image Understanding task and the manipulation reasoning questions in the Image Reasoning task, five licensed acupuncturists were hired to construct single-choice questions based on images and their associated textual descriptions. The constructed items were reviewed and revised by two additional acupuncturists to ensure quality and accuracy.

For the acupoint identification subtask within the Image Understanding task, let the set of acupoints annotated in a given image $I$ be defined as $\mathcal{L}_t = \{a_1, \ldots, a_n\}$, representing the acupoints actually depicted in the image. Let $c(\cdot)$ denote a mapping function from individual acupoints to their corresponding acupoint categories, and let $\mathcal{C}_k$ denote the complete set of acupoints under the $k$-th category. Based on this, the distractor candidate set can be defined as:
\[
\mathcal{L}_f =  \{x \mid x \in \bigcup_{a_i \in \mathcal{L}_t} \mathcal{C}_{c(a_i)}, \quad x \notin \mathcal{L}_t \}
\]


which includes all acupoints belonging to the same category as those in the image but not present in the image itself.

\paragraph{Single-choice Question Construction:}
For single-choice question construction, each acupoint $a_k \in \mathcal{L}_t$ is selected as the sole correct answer, and an ordered, non-repeating sequence $(b_1, b_2, b_3)$ is sampled from $\mathcal{L}_f$, where $b_i \in \mathcal{L}_f$ and $b_i \neq b_j$ for $i \neq j$. The element order is determined by a predefined index or traversal rule. This yields a candidate option set $
\{a_k, b_1, b_2, b_3\}$.

\paragraph{Multiple-choice Question Construction:}
For multiple-choice question construction, a subset of size $m$ (where $2 \leq m \leq 4$) is sampled from $\mathcal{L}_t$ as the correct answer set. Additionally, $n$ distractors (where $0 \leq n \leq 2$) are drawn from $\mathcal{L}_f$ using the same ordered, non-repeating sampling strategy.

\subsubsection{Data Statistics}
In this study, a total of 10,729 visual question answering (VQA) items were constructed. These items cover 12 acupoint categories (including the Hand Jueyin Meridian) and encompass 304 unique acupoints represented by typical examples such as Baihui (GV20). The overall statistical information of the curated dataset is illustrated in Figure \ref{fig:vqa-statistics}, while the data distribution of the 12 acupoint categories in the validation set is detailed in Table~\ref{tab:app-vqa-distribution} in the Appendix. Finally, each data instance in the Hand Taiyin Lung Meridian category was visualized, as illustrated in Figure \ref{fig:vqa-statistics}.

\subsection{AcupunctureVideo}
\begin{figure*}[!t]
\centering
\includegraphics[width=\textwidth]{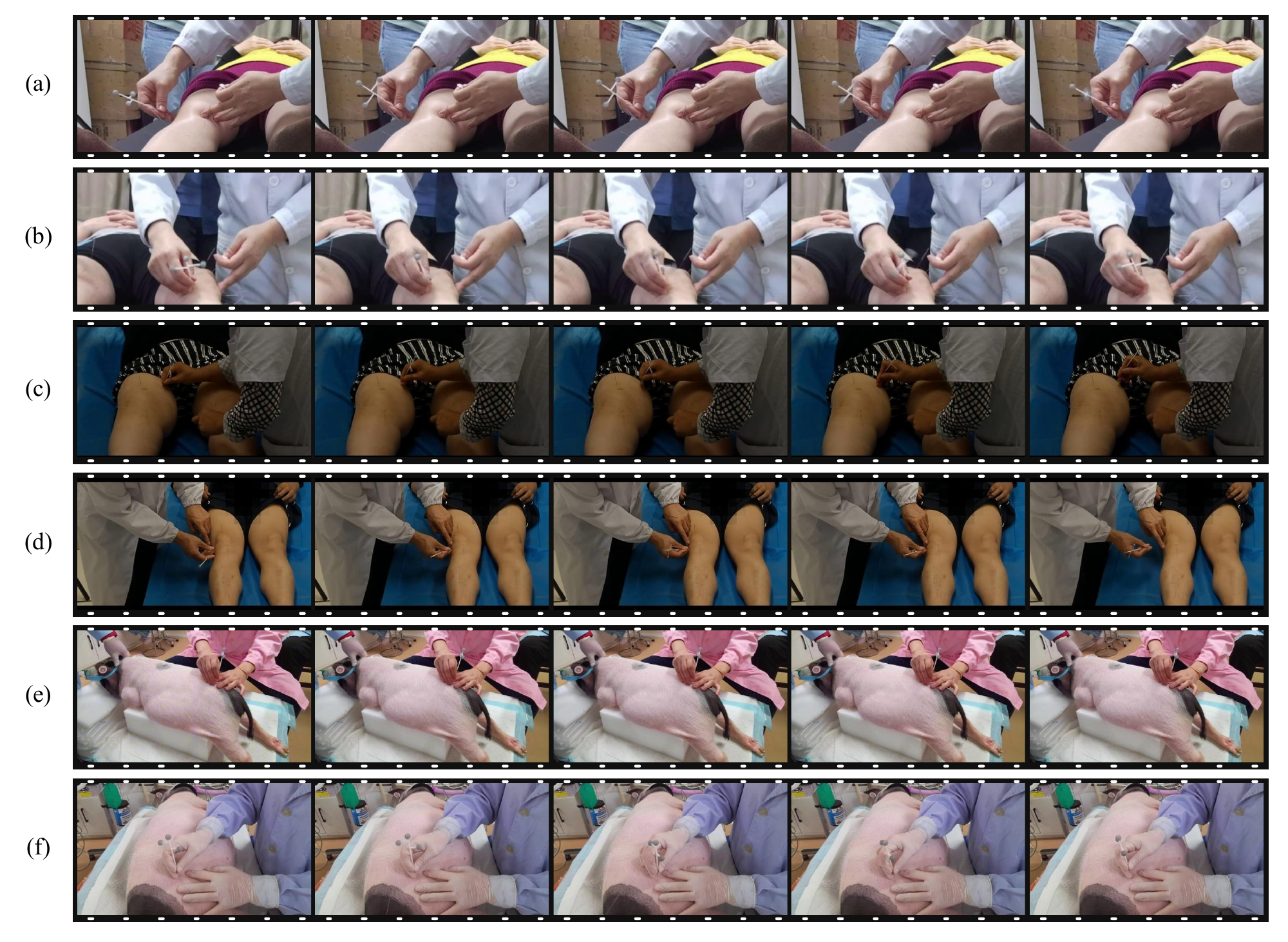} 
\caption{ Visualization of three categories of \textbf{AcupunctureVideo}. 
(a)–(b): Category 1, human Dubi (ST 35) and Neixi acupoint manipulation; (c)–(d): Category 2, clinical manipulations by different acupuncturists on distinct patients; (e)–(f): Category 3, porcine Baihui (GV 20) and Xiaokua acupoint manipulation.
}
\label{fig:video-overview}
\end{figure*}
A total of 1,216 acupuncture practice videos were collected to construct the Acupuncture Video Datasets, illustrated in Figure \ref{fig:video-overview}. The video content was systematically divided into three major categories, with detailed statistics presented in Figure~\ref{fig:app-video-statistics} of the Appendix. The first category comprises 300 standardized practice videos, capturing 15 acupuncturists performing needling techniques on 20 commonly used acupoints. These videos aim to document the procedural standards and to reflect variation in technique across different practitioners for the same acupoints. The second category includes 458 videos recorded in real clinical settings, featuring six acupuncture experts administering treatments to patients. These videos closely represent real-world clinical scenarios and enhance the dataset’s ecological validity and diversity. The third category consists of 458 experimental videos capturing acupoint manipulation on pigs. This segment focuses on five target acupoints and serves as foundational data for research into acupuncture applications in animal models.


\subsubsection{Video Acquisition Process}
Acupuncture manipulation videos were collected using an optical motion capture system, as illustrated in Fig.~\ref{fig:app-acquisition} of the Appendix. Multiple synchronized cameras were employed to operate collaboratively during the recording process. This setup significantly enhanced adaptability to clinical environments, enabling the unobtrusive capture of needling techniques with minimal disruption to both practitioners and patients. The system is well-suited for complex clinical settings and effectively reduces external interference during acupuncture procedures. During the data acquisition and preprocessing stages, all videos underwent a quality control process to remove samples containing excessive noise or irrelevant content.  Furthermore, privacy protection measures were applied to all videos.  The entire data collection protocol was approved by the relevant ethics committee, and authorization details are provided.  The hardware specifications used for video capture are illustrated in Fig.~\ref{fig:app-acquisition} of the Appendix.

\section{Experiments}
\label{sec:experiments}

\begin{table*}[!t]

\centering

\setlength{\tabcolsep}{2pt}  
\begin{adjustbox}{max width=\textwidth}
\begin{tabular}{cccccccc}
\hline

\multirow{1}{*}{\textbf{Class}} & \multirow{1}{*}{\textbf{Models}} & \multicolumn{1}{c}{\textbf{Consistency}} &\multicolumn{1}{c}{\textbf{Accuracy}} &\multicolumn{1}{c}{\textbf{Logic}} &\multicolumn{1}{c}{\textbf{Fluency}} &\multicolumn{1}{c}{\textbf{Security}} &\multicolumn{1}{c}{\textbf{Sum.}}\\
\hline

\multirow{7}{*}{\makecell{Domain-specific \\ LLMs}}
&Zhongjing-LLaMA-base & 3.30 & 14.81 & 2.68 & 5.74 & 5.72 & 32.25 \\
&Lingdan-13B-PR & 3.26 & 17.18 & 2.68 & 5.14 & 5.65 & 33.91 \\
&HuatuoGPT2-7B & 6.36 & 25.21 & 5.05 & 8.48 & 6.81 & 51.91 \\
&HuatuoGPT2-13B & 6.23 & 26.85 & 5.23 & 8.65 & 6.81 & 53.77 \\
&BianCang-Qwen2.5-7B-Instruct & 6.69 & 29.93 & 5.41 & 7.09 & 7.02 & 56.14 \\

&HuatuoGPT2-34B & 6.96 & 32.81 & 6.15 & 8.77 & 7.68 & 62.37 \\
&HuatuoGPT-o1-72B & 9.53 & 49.06 & 8.96 & 9.74 & 9.56 & 86.85 \\
 \hline
 \multirow{7}{*}{\makecell{General-purpose \\ LLMs}} 
 &gpt-4o-mini & 8.27 & 31.38 & 6.35 & 9.03 & 7.61 & 62.64 \\
&gpt-4.1-mini & 9.28 & 41.02 & 8.27 & 9.44 & 8.92 & 76.93 \\
&ERNIE-4.5-Turbo-128K & 8.35 & 47.88 & 7.88 & 8.98 & 9.29 & 82.38 \\
&DeepSeek-R1-250528 & 8.64 & 50.05 & 8.24 & 8.99 & 9.20 & 85.12 \\
&ERNIE-X1-Turbo-32K & 9.04 & 52.70 & 8.50 & 9.06 & 9.50 & 88.80 \\
&Kimi-K2-Instruct & 9.14 & 54.00 & 8.80 & 9.04 & 9.56 & 90.54 \\

&gpt-5-mini & 9.88 & 56.47 & 9.67 & 9.92 & 9.85 & 95.79 \\
\hline
\end{tabular}
\end{adjustbox}
\caption{
Performance of LLMs on the open-ended subjective question answering test set in the acupuncture domain. “Sum.” denotes the total score across all dimensions, with scores ranging from 0 to 100.
  }
  \label{tab:qa-performance}
\end{table*}

\begin{table}[htbp]
\centering

\begin{adjustbox}{max width=\columnwidth}
\begin{tabular}{llll}
\hline
LLMs                    & $r$    & $\rho$ & $\tau$ \\ \hline
Kimi-K2-Instruct        & 0.832  & 0.816  & 0.653  \\ 
DeepSeek-R1-250528      & 0.805  & 0.781  & 0.617  \\ \hline
\end{tabular}
\end{adjustbox}
\caption{Correlations between ERNIE-X1-Turbo-32K and licensed acupuncturists on the open-ended subjective QA test set in the acupuncture domain.}
\label{tab:human-correlation}
\end{table}

\begin{table*}[!t]

\centering
\begin{adjustbox}{max width=\textwidth}
\begin{tabular}{ccccccccc}
\hline

\multirow{1}{*}{\textbf{Class}} & \multirow{1}{*}{\textbf{Models}} &\multicolumn{1}{c}{\textbf{A1}} &\multicolumn{1}{c}{\textbf{A2}} &\multicolumn{1}{c}{\textbf{A3}} &\multicolumn{1}{c}{\textbf{A4}} &\multicolumn{1}{c}{\textbf{B}} &\multicolumn{1}{c}{\textbf{X}} &\multicolumn{1}{c}{\textbf{Avg.}}\\
\hline
\multirow{7}{*}{\makecell{Domain-specific \\ LLMs}}
& Lingdan-13B-PR & 14.29 & 25.64 & 0.00 & 0.00 & 15.00 & 25.00 & 13.32 \\
& HuatuoGPT2-13B & 17.86 & 17.95 & 15.00 & 25.00 & 7.50 & 20.00 & 17.22 \\
& Zhongjing-LLaMA-base & 23.21 & 20.51 & 12.50 & 20.83 & 10.00 & 25.00 & 18.68 \\
& HuatuoGPT2-7B & 19.64 & 30.77 & 35.00 & 20.83 & 20.05 & 15.00 & 23.55 \\
& HuatuoGPT2-34B & 26.79 & 41.03 & 37.50 & 25.00 & 17.50 & 20.00 & 27.97 \\
& BianCang-Qwen2.5-7B-Instruct & 58.93 & 87.18 & 50.00 & 54.17 & 62.50 & 25.00 & 56.30 \\
& HuatuoGPT-o1-72B & 73.21 & 89.74 & 85.00 & 70.83 & 80.00 & 35.00 & 72.30 \\

\hline
\multirow{7}{*}{\makecell{General-purpose \\ LLMs}} 
& gpt-4.1-mini & 37.50 & 53.85 & 70.00 & 62.50 & 55.00 & 20.00 & 49.81 \\
& gpt-4o-mini & 35.71 & 58.97 & 70.00 & 58.33 & 55.00 & 25.00 & 50.50 \\
& gpt-5-mini & 51.79 & 79.49 & 82.50 & 66.67 & 65.00 & 50.00 & 65.91 \\
& ERNIE-4.5-Turbo-128K & 73.21 & 92.31 & 85.00 & 83.33 & 85.00 & 45.00 & 77.31 \\
& ERNIE-X1-Turbo-32K & 85.71 & 94.87 & 87.50 & 87.50 & 85.00 & 60.00 & 83.43 \\
& Kimi-K2-Instruct & 80.36 & 87.18 & 92.50 & 83.33 & 90.00 & 70.00 & 83.90 \\
& DeepSeek-R1-250528 & 89.29 & 92.31 & 90.00 & 83.33 & 85.00 & 65.00 & 84.16 \\
\hline

\end{tabular}
\end{adjustbox}
\caption{
Performance of LLMs on the AcupunctureExam test set. Each alphanumeric code (e.g., A1) represents a distinct question category. “Avg.” indicates the mean score across all categories, on a 0–100 scale.}
\label{tab:exam-performance}
\end{table*}

\subsection{Evaluated Generative Models}
We evaluated leading text-only large language models, large multimodal models, and video foundation models on the AcupunctureBench benchmark, encompassing both domain-specific models and general-purpose models.

\subsubsection{Large Language Models}

\textbf{Domain-specific LLMs:} Lingdan-13B-PR\cite{hua2024lingdan}, Zhongjing-LLaMA-base\cite{yang2024zhongjing}, HuatuoGPT2-7B\cite{chen2023huatuogpt}, HuatuoGPT2-13B\cite{chen2023huatuogpt}, HuatuoGPT2-34B\cite{chen2023huatuogpt}, HuatuoGPT-o1-72B\cite{chen2024huatuogpt}, BianCang-Qwen2.5-7B-Instruct\cite{wei2025biancang}.
\noindent\textbf{General-purpose LLMs:} gpt-4o-mini\cite{hurst2024gpt}, gpt-4.1-mini\footnote{\url{https://openai.com/gpt-4/}}, gpt-5-mini\footnote{\label{ft:gpt5}\url{https://openai.com/gpt-5/}}, ERNIE-4.5-Turbo-128K\cite{sun2019ernie}, ERNIE-X1-Turbo-32K\cite{sun2019ernie}, Kimi-K2-Instruct\cite{team2025kimi}, DeepSeek-R1-250528\cite{guo2025deepseek}. \textbf{Large Multimodal Models}
gpt-5-nano\reusefootnote{ft:gpt5}, gpt-5.2\reusefootnote{ft:gpt5}, ERNIE-4.5-VL-28B-A3B\cite{sun2019ernie}, ERNIE-4.5-Turbo-VL-32K\cite{sun2019ernie}, InternVL2-26B\cite{chen2024expanding}, InternVL2.5-8B\cite{chen2024expanding}, grok-4-fast\footnote{\url{https://x.ai/news/grok-4-fast}}, grok-4.1-fast\footnote{\url{https://x.ai/news/grok-4-1}}, claude-haiku-4.5\footnote{\url{https://www.anthropic.com/news/claude-haiku-4-5}}, Qwen2.5-VL-7B-Instruct\cite{bai2025qwen2}, Qwen3-VL-235B-A22B-Instruct\footnote{\url{https://github.com/QwenLM/Qwen3-VL}}, glm-4.6v\cite{hong2025glm}, gemini-2.5-flash\cite{comanici2025gemini}, gemini-3-flash-preview\footnote{\url{https://blog.google/products/gemini/gemini-3-flash/}}.
\textbf{Video Foundation Models}
nemotron-nano-12b-v2-vl\cite{deshmukh2025nvidia}, glm-4.6v\cite{hong2025glm}, gemini-2.5-flash-lite-preview-09-2025\cite{comanici2025gemini}, gemini-3-flash-preview\footnote{\url{https://blog.google/products/gemini/gemini-3-flash/}}.

\begin{table*}[!t]

\centering
\begin{adjustbox}{max width=\textwidth}
\begin{tabular}{cccccc}
\hline

\multirow{1}{*}{\textbf{Models}} &\multicolumn{1}{c}{\textbf{Type1(s)}} &\multicolumn{1}{c}{\textbf{Type1(m)}} &\multicolumn{1}{c}{\textbf{Type2}} &\multicolumn{1}{c}{\textbf{Type3}} &\multicolumn{1}{c}{\textbf{Avg.}}\\
\hline
gpt-5-nano & 40.12 & 10.67 & 70.27 & 45.71 & 41.69 \\
ERNIE-4.5-VL-28B-A3B & 24.38 & 20.33 & 72.97 & 54.29 & 42.99 \\
InternVL2-26B & 66.98 & 24.00 & 37.84 & 57.14 & 46.49 \\
Qwen2.5-VL-7B-Instruct & 52.47 & 29.83 & 75.68 & 54.29 & 53.07 \\
InternVL2.5-8B & 59.88 & 20.50 & 81.08 & 65.71 & 56.79 \\
grok-4-fast & 59.57 & 39.33 & 75.68 & 57.14 & 57.93 \\
grok-4.1-fast & 58.02 & 39.33 & 78.38 & 68.57 & 61.08 \\

claude-haiku-4.5 & 61.73 & 61.67 & 72.97 & 62.86 & 64.81 \\

gemini-2.5-flash & 90.43 & 66.17 & 67.57 & 54.29 & 69.62 \\
ERNIE-4.5-Turbo-VL-32K & 70.68 & 65.67 & 91.89 & 54.29 & 70.63 \\
Qwen3-VL-235B-A22B-Instruct & 88.89 & 77.33 & 91.89 & 71.43 & 82.39 \\
glm-4.6v & 89.51 & 78.17 & 89.19 & 71.43 & 82.08 \\
gpt-5.2 & 92.90 & 81.00 & 91.89 & 82.86 & 87.16 \\
gemini-3-flash-preview & 93.83 & 89.83 & 94.59 & 88.57 & 91.71 \\
\hline
\end{tabular}
\end{adjustbox}
\caption{
Performance of LMMs on the AcupunctureVQA test set. Type 1 and Type 2 correspond to acupoint identification and acupoint localization tasks in Image Understanding, respectively, where (s) denotes single-choice questions and (m) denotes multiple-choice questions. Type 3 refers to manipulation reasoning questions in Image Reasoning. “Avg.” denotes the average score across all question types, with scores ranging from 0 to 100.
  }
\label{tab:vqa-performance}
\end{table*}

\subsection{Experimental Results}
\textbf{Performance of LLMs on the AcupunctureQA Test Set:} We firstly evaluated the performance of LLMs on open-ended subjective question answering in the acupuncture domain, as shown in Table \ref{tab:qa-performance}.  Among the domain-specific LLMs, only HuatuoGPT2-34B and HuatuoGPT-o1-72B exceeded the pass threshold and the good threshold, achieving scores of 62.37 and 86.85, respectively.  Among the general-purpose LLMs, all models surpassed the pass threshold.  Both Kimi-K2-Instruct and gpt-5-mini exceeded the excellent threshold.  To quantitatively assess the consistency between ERNIE-X1-Turbo-32K and the human evaluators, we additionally recruited three licensed acupuncturists to independently score the responses generated by two general-purpose LLMs on the expanded open-ended subjective QA test set in the acupuncture domain.   Meanwhile, we employed three statistical coefficients: Pearson’s correlation coefficient ($r$), Spearman’s rank correlation coefficient ($\rho$), and Kendall’s tau coefficient ($\tau$). These coefficients served as meta-evaluation metrics for comparing the correlation between ERNIE-X1-Turbo-32K and the acupuncturists’ ratings.  The experimental results are presented in Table \ref{tab:human-correlation}.  The results indicate that the model’s evaluation scores exhibit a high level of correlation with the ratings provided by the acupuncturists.

Finally, we evaluated LLM performance on the AcupunctureExam benchmark, as shown in Table \ref{tab:exam-performance}.  Among the domain-specific LLMs, only HuatuoGPT-o1-72B exceeded the pass threshold, achieving a score of 72.30.  None of the general-purpose LLMs reached the excellent threshold;  the highest score was 84.16, achieved by DeepSeek-R1-250528.  Notably, the GPT-series models exhibited the weakest performance among the general-purpose LLMs, with only gpt-5-mini reaching the pass threshold.

\begin{figure}[htbp]
\centering
\includegraphics[width=\columnwidth]{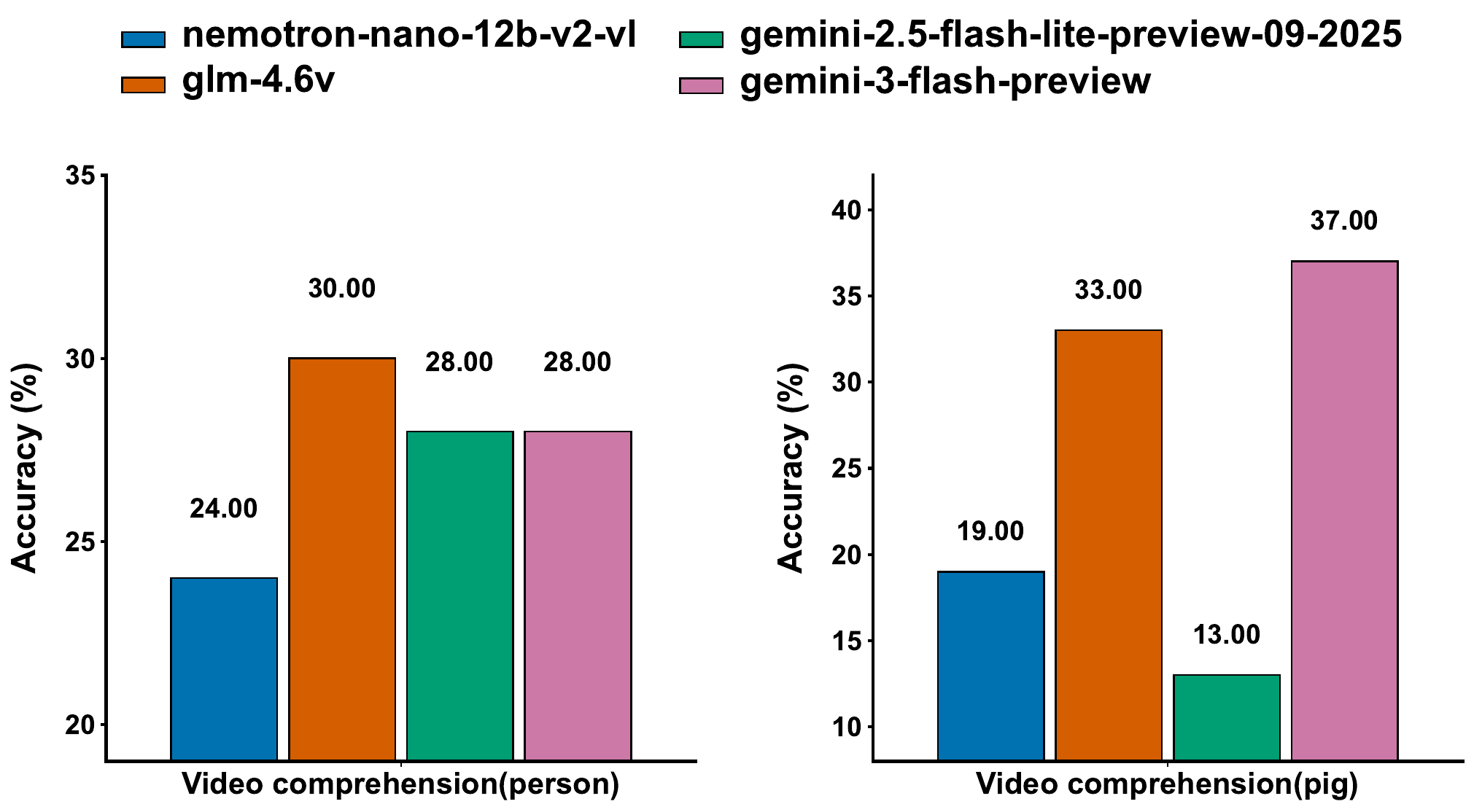} 
\caption{Performance of ViFMs on the AcupunctureVideo test set.}
\label{fig:video-performance}
\end{figure}

\noindent\textbf{Performance of LMMs on the Acupuncture VQA Test Set:} We evaluated state-of-the-art LMMs on AcupunctureVQA in the acupuncture domain, with the results summarized in Table \ref{tab:vqa-performance}. Overall, 57.14\% and 28.57\% of the evaluated LMMs exceeded the pass threshold and the good threshold, respectively. Only gemini-3-flash-preview surpassed the excellent threshold, outperforming the second-ranked model, gpt-5.2, by 4.55 points. We observed that the largest performance gaps occurred on Type 1 (m) and Type 3 question categories, indicating that gemini-3-flash-preview demonstrates superior capabilities in complex image understanding and multimodal reasoning compared to gpt-5.2 in the acupuncture domain.

\noindent \textbf{Performance of ViFMs on the AcupunctureVideo Test Set:} Finally, we evaluated four state-of-the-art ViFMs on the AcupunctureVideo video-based judgment task, as shown in Fig. \ref{fig:video-performance}. The findings show that state-of-the-art ViFMs perform poorly on identifying basic acupuncture manipulation techniques.  The best-performing model, gemini-3-flash-preview, achieved an accuracy of only 37 percent. These results objectively indicate that ViFMs still have substantial room for improvement on video comprehension tasks involving acupuncture point localization and manipulation techniques.

\section{Conclusion}
\label{sec:conclusion}
In this work, we present AcupunctureBench, the first large-scale multimodal benchmark for the acupuncture domain. By integrating text, images, and videos, it addresses a critical gap in domain-specific multimodal evaluation. All question–answer pairs are independently constructed from authoritative textbooks, avoiding reuse of public datasets and reducing data leakage. We introduce an efficient deduplication strategy to support scalable data construction. We also release the first dedicated evaluation website for acupuncture, enabling transparent and standardized assessment. We hope this work will inspire future research on foundation models for TCM domain.

\section*{Limitations}

As the first large-scale multimodal evaluation benchmark for acupuncture, AcupunctureEval is exploratory but has limitations. It does not develop or compare a domain-specific acupuncture model to avoid potential bias from evaluating a self-built model on a custom benchmark. The data are mainly drawn from Chinese textbooks, with limited international literature and clinical experience. Image and video samples are scarce, and real clinical scenarios are insufficiently covered. Moreover, the framework focuses on understanding and reasoning, without systematic evaluation of safety or interpretability.

\bibliography{acupunctureeval}
\clearpage

\appendix
\renewcommand{\theHfigure}{appendix.\arabic{figure}}
\renewcommand{\theHtable}{appendix.\arabic{table}}
\setcounter{figure}{0} 
\renewcommand{\thefigure}{A\arabic{figure}} 
\setcounter{table}{0}
\renewcommand{\thetable}{A\arabic{table}}

\phantomsection
\label{app:start}
\textbf{Appendix}
\section{Related Work}
\label{app:related-work}

\subsection{Benchmarks in the TCM domain}
Inspired by benchmark studies in the general medical domain, the field of TCM has gradually initiated evaluation explorations for LLMs and proposed a series of text based benchmarks, including TcmBench\cite{yue2024tcmbench}, TCMD\cite{yu2024tcmd}, TCMEval-SDT\cite{wang2025tcmeval}, TCM-3CEval\cite{huang2026triaxial}, TCM-5CEval\cite{huang2025tcm}, DCB\cite{shi2026benchmarking}, and MTCMB\cite{kong2025mtcmb}. These studies mainly focus on traditional Chinese medicine theoretical knowledge, diagnostic and therapeutic question answering, and reasoning abilities, and they evaluate model text generation performance within traditional Chinese medicine contexts.  

\subsection{General medical domain benchmarks}
Early research on benchmark evaluation of LLMs in the medical domain mainly focused on Western medicine settings and relied on English medical corpora, leading to the construction of several representative evaluation frameworks such as MMLU\cite{hendryckstest2021,hendrycks2021ethics}, MedQA\cite{jin2021disease}, MedMCQA\cite{pal2022medmcqa}, and MultiMedQA\cite{singhal2023large}. These benchmarks typically center on tasks including medical question answering, clinical reasoning, and medical knowledge understanding, and they provide systematic assessments of models with respect to medical knowledge acquisition and reasoning capability. On this basis, researchers gradually began to pay attention to Chinese medical scenarios and successively developed multiple benchmark evaluation systems oriented toward Chinese medical corpora, such as CBLUE\cite{zhang2022cblue}, ExplainCPE\cite{li-etal-2023-explaincpe}, MedEval\cite{he2023medeval}, CMB\cite{wang2024cmb}, and MedBench\cite{cai2024medbench}  effectively alleviating the shortage of evaluation resources in Chinese medical language processing.

\begin{figure*}[!t]
\centering
\includegraphics[width=\textwidth]{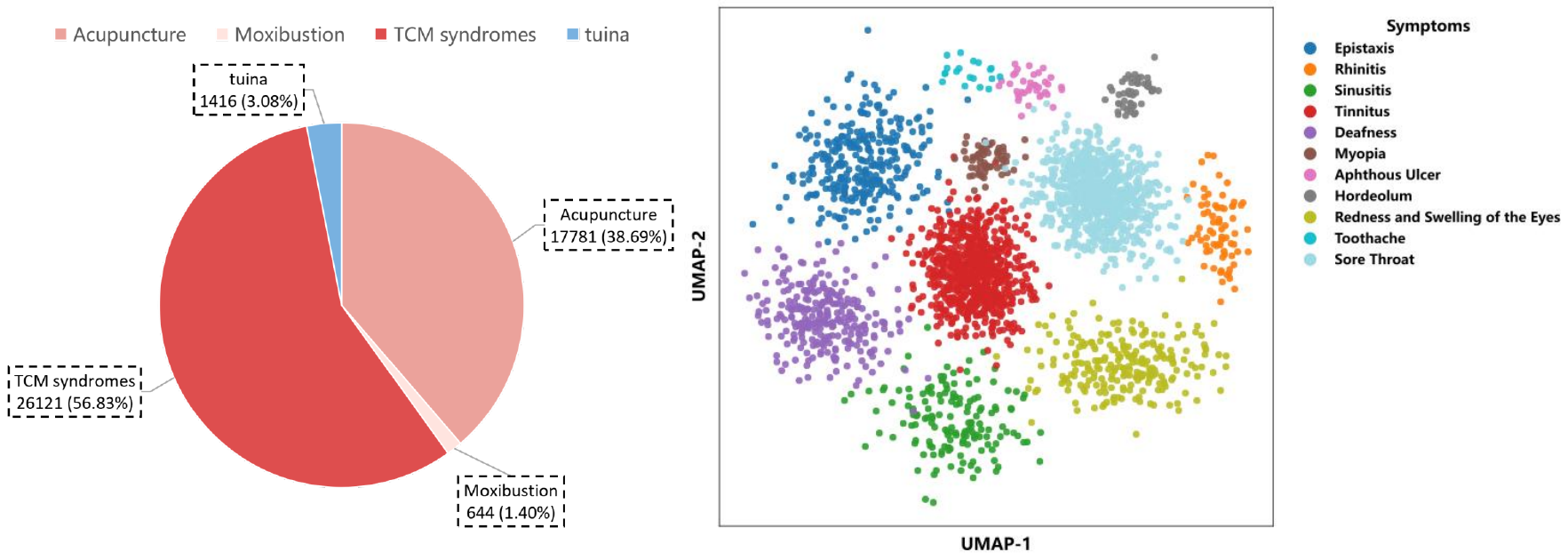} 
\caption{ The left panel shows the visualization of data statistics for open-ended subjective question answering, while the right panel presents the visualization of individual items associated with syndromes belonging to the otorhinolaryngology and sensory organ category under the traditional Chinese medicine (TCM) syndromes domain in the open-ended subjective question answering dataset.
}
\label{fig:app-qa-statistics}
\end{figure*}

\subsection{Evaluation approaches in the acupuncture domain}
Compared with other branches of traditional Chinese medicine, systematic evaluation research on LLMs in the acupuncture domain remains more limited.  Existing studies mostly approach this topic from the perspective of evaluation methods or analytical frameworks.  For example, the Sentence Similarity Evaluation Module evaluates model\cite{li2025taming} performance on acupuncture related tasks through sentence similarity computation, and other study\cite{liu2025evaluating} has proposed evaluation frameworks to compare differences between LLMs and acupuncturists in clinical case responses. Although these studies provide methodological references for evaluating LLMs in the acupuncture domain, they primarily focus on evaluation strategies themselves and have not yet established a benchmark specifically designed for acupuncture.

    
    
    

\section{Statement}
\label{app:statement}
\subsection{Leakage Prevention Statement}
Evaluation data must be used exclusively for model or human medical assessment. It must not be used for training in any capacity, including the training of synthetic or modified versions of the data.



\subsection{Ethics Statement}
\textbf{Human Studies}
This study was approved by the Ethics Committee of XXX Hospital (Approval No. XXX). Written informed consent was obtained from all participants. All procedures were conducted in accordance with the principles of the Declaration of Helsinki.

\noindent\textbf{Animal Experiments} All animal experiments were approved by the Institutional Animal Care and Use Committee (IACUC) of XXX Hospital (Approval No.XXX) and were performed in compliance with the ARRIVE guidelines.

\subsection{LLM Usage Statement}

Large Language Models (LLMs) were employed to assist in polishing the manuscript. Specifically, we utilized LLM support to optimize linguistic expression. Tasks completed with the model’s assistance included: sentence rewriting, grammar checking, and enhancing overall textual coherence. It is important to note that the LLM did not participate in the conception of the research, the design of the methodology, or the formulation of the experimental plan. All research concepts, ideas, and analyses were independently developed by the authors. The authors bear full responsibility for the content of the manuscript, including text generated or polished by the LLM. We have ensured that text generated by the LLM complies with ethical guidelines and does not constitute plagiarism or scientific misconduct.


\section{Appendix to the Method Section}
\label{app:methods}
\subsection{Evaluation Protocol}
First, different evaluation strategies were adopted for each sub-dataset in AcupunctureQA. Specifically, for the test set of AcupunctureExam, top-1 prediction accuracy was\cite{krizhevsky2012imagenet} employed as the evaluation metric. 

tinyBenchmarks\cite{polo2024tinybenchmarks} indicates that, under the optimal IRT++ evaluation strategy, just 100 carefully selected samples from each scenario are sufficient to keep the average estimation error of LLM performance within 2\% across the four major benchmarks. Therefore, we randomly sampled 100 items from each of the two domains, Acupuncture and TCM syndromes, to form the evaluation set. A prompt template, presented in \hyperref[app:qa-evaluation]{Subsection II} of Section~\ref{app:prompts}, was designed to guide ERNIE-X1-Turbo-32K\cite{sun2019ernie} to produce evaluation outputs along five predefined dimensions. Three licensed acupuncturists were recruited to score the responses in AcupunctureClin according to the evaluation criteria detailed in \hyperref[app:clin-rubric]{Subsection III} of Section~\ref{app:prompts}. 

For the test set of AcupunctureVQA, top-1 prediction accuracy\cite{krizhevsky2012imagenet} was also used as the evaluation metric for single-choice questions. For multiple-choice questions, we adopted a more stringent evaluation protocol: an item was considered correct only if the model selected all correct options; otherwise, a score of zero was assigned. 

Finally, for the AcupunctureVideo test set, top-1 prediction accuracy\cite{krizhevsky2012imagenet} was likewise used as the evaluation metric. Across all tasks, we adopted the widely used examination cutoffs to benchmark model performance: 60 as the pass threshold, 80 as the good threshold, and 90 as the excellent threshold.

\subsection{Data Sources}
\subsubsection{AcupunctureQA Data Sources}
A total of 28 classical texts and authoritative textbooks on acupuncture, along with relevant entries from medical encyclopedias\footnote{\url{https://zh.wikipedia.org/wiki}}, were selected as the original data sources for constructing the QA dataset. This approach was intended to ensure data quality from the source. Different processing strategies were employed based on the document type. For PDF files containing extractable text layers, the PyMuPDF\footnote{\url{https://github.com/pymupdf/PyMuPDF}} was used to parse and extract textual content. For scanned PDFs, the GLM-4.6v-flash\cite{hong2025glm} was first utilized to perform OCR on the image content, converting it into editable plain text. Subsequently, regular expressions were uniformly applied to the extracted text to segment it at the paragraph level, thereby constructing a repository of candidate textual units.

\subsubsection{AcupunctureVQA Data Sources}
We encountered two key challenges in acquiring acupuncture-related images from the internet: the overall image quality is often low, and the available image data are limited in quantity. To address these issues, we utilized images and their corresponding textual descriptions from the higher education textbook \emph{Meridians and Acupoints}, which served as a high-quality data source for the acupuncture domain.

\subsection{AcupunctureExam}
\subsubsection{Data Sources}
In this study, the construction of the objective-question component did not incorporate data sources derived from the internet or existing question banks, which are commonly used in most current benchmarks. This decision was motivated by the absence of a standardized question bank in the field of acupuncture. Moreover, acupuncture-related questions available online are highly fragmented, making it difficult to systematically collect samples and ensure their accuracy. More importantly, we observed that most existing benchmarks emphasize the questions themselves while neglecting a fine-grained categorization of question types. We argue that, for LLM evaluation tasks, a detailed, task-oriented analysis focusing on individual sub-tasks is more effective in revealing model strengths and weaknesses than evaluating performance based solely on large-scale aggregated datasets. To address these issues, the objective-question dataset used in this study was exclusively derived from the multiple-choice questions in the textbook \emph{Collection of Exercises for Acupuncture Studies}.

\subsubsection{Data Statistics}
\begin{figure}[H]
\centering
\includegraphics[width=\columnwidth]{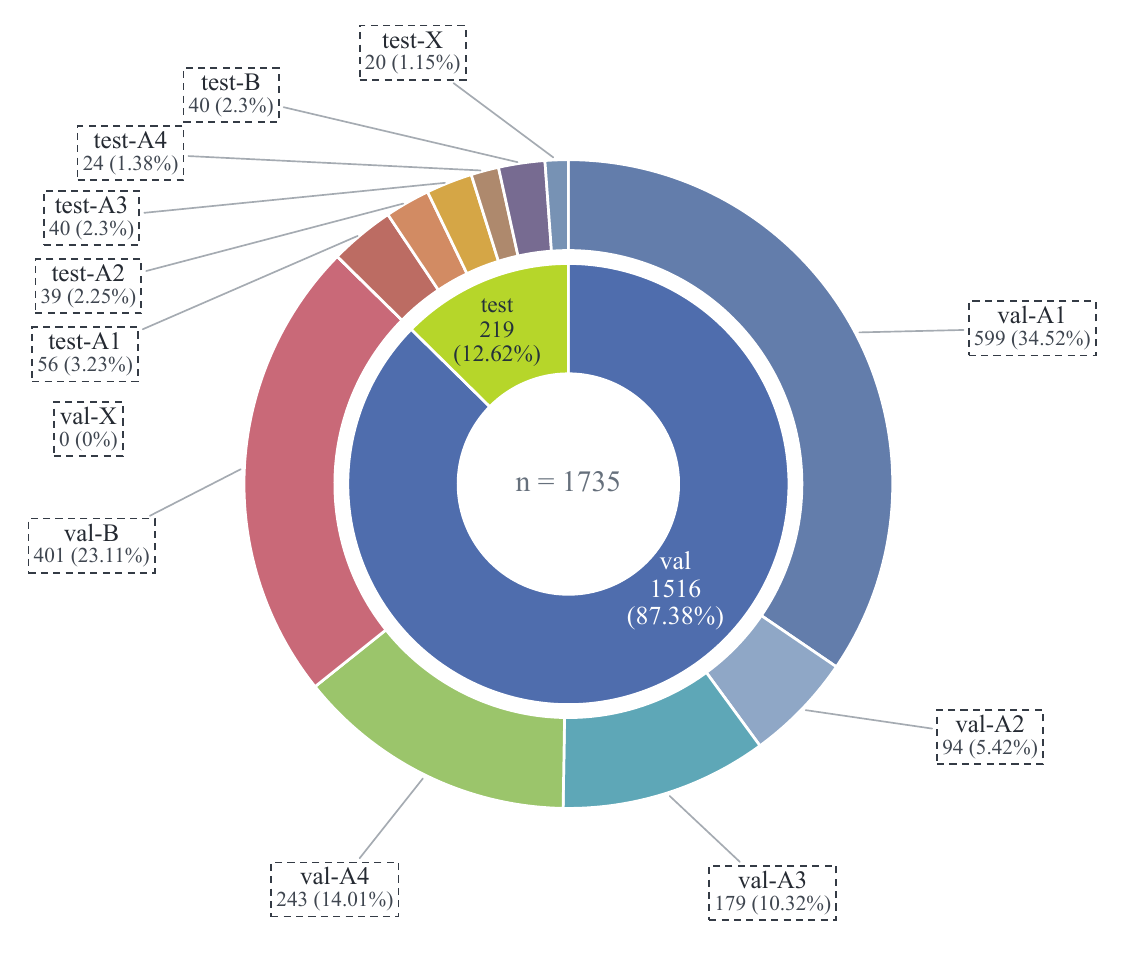} 
\caption{Visualization of AcupunctureExam Data
Statistics.}
\label{fig:app-exam-statistics}
\end{figure}

The multiple-choice section consists of Type A, Type B, and Type X questions. Type A questions are further subdivided into A1, A2, A3, and A4 categories, where A1 questions are descriptive in nature, and A2, A3, and A4 questions are case-based. Type B questions correspond to standard compatibility questions, while Type X questions are descriptive multiple-answer questions. The objective-question dataset contains a total of 1,516 samples, covering three major categories: meridians and acupoints, acupuncture and moxibustion techniques, and acupuncture-related diseases. Among these, the meridians and acupoints category includes 15 subcategories, while the acupuncture-related diseases category spans four major clinical departments and encompasses 78 distinct diseases. In addition, 219 questions were selected as the test set. To mitigate potential data contamination, the order of answer options was randomly perturbed, and users are required to upload their self-evaluation results to our publicly available evaluation platform for scoring. A statistical overview of the objective-question dataset is presented in Fig. \ref{fig:app-exam-statistics}.

\subsection{AcupunctureClin}
To evaluate the performance of LLMs in acupuncture-related clinical settings, we selected 100 clinical case questions from the textbook \emph{Collection of Exercises for Acupuncture Studies}. Each case was assessed across six dimensions to comprehensively measure LLMs performance: Pathogenesis, Diagnosis, Subtyping, Treatment  Protocol, Acupoint Selection, and Rationale.

\section{Appendix to the Experimental Section}
\label{app:experiments}
We also evaluated LLMs on open-ended, subjective question answering in the TCM syndromes domain, with results summarized in Table~\ref{tab:app-syndromes-performance} of the Appendix. Among the domain-specific LLMs, all models in the HuatuoGPT2 series as well as BianCang-Qwen2.5-7B-Instruct exceeded the pass threshold, with HuatuoGPT-o1-72B achieving the highest score of 86.22. Among the general-purpose LLMs, all models surpassed the good threshold. gpt-5-mini exceeded the excellent threshold.

Three licensed acupuncturists evaluated LLM performance on AcupunctureClin according to predefined scoring criteria. The evaluation results are shown in Table~\ref{tab:app-clin-performance} of the Appendix, and the scoring rubric is presented in the Appendix. Among the domain-specific LLMs, Lingdan-13B-PR and HuatuoGPT-o1-72B exceeded the pass threshold, achieving scores of 62.35 and 74.71, respectively. Among the general-purpose LLMs, all models also surpassed the pass threshold; however, none exceeded the excellent threshold. ERNIE-X1-Turbo-32K achieved the highest score of 89.41.

Initially, due to computational resource constraints and considerations regarding model response time, we limited our evaluation to two key aspects within the field of acupuncture: acupuncture techniques and TCM syndromes. We sought to understand how LLMs perform when the validation set is expanded. To this end, we randomly selected an additional 10\% of the questions to serve as a new evaluation sample. This increased the test sample by 1,779 entries. The experimental results are shown in Table Table~\ref{tab:aug}. The results indicate that the conclusions drawn from the expanded test set are largely consistent with those based on the original dataset. The main difference is that, after increasing the sample size, the model’s overall accuracy decreased slightly, by approximately 2\%.



\begin{table*}[!t]
  \centering

\begin{adjustbox}{max width=\textwidth}
  \begin{tabular}{c l} 
    \toprule
    Dimension & Description \\
    \midrule
    Clinical Accuracy & 
    \makecell[l]{
      0: Presence of obvious errors or pseudoscientific content \\ 
      1: Generally correct but with imprecise phrasing \\ 
      2: Correct and precise use of medical terminology
    } \\ \hline
    
    Question Clarity & 
    \makecell[l]{
      0: Vague question or severe grammatical errors \\ 
      1: Semantics generally clear but non-standard \\ 
      2: Professional wording with a clear structure
    } \\ \hline
    
    Answer Completeness & 
    \makecell[l]{
      0: Only partially addresses the question; key points missing \\ 
      1: Main points covered but with slight omissions or logical gaps \\ 
      2: Complete answer with clear structure; causes and effects explained
    } \\ \hline
    
    Terminology Consistency & 
    \makecell[l]{
      0: Confusing or incorrect use of terminology \\ 
      1: Generally accurate but inconsistent in terminology \\ 
      2: Rigorous adherence to TCM standards
    } \\ \hline
    
    Clinical Utility & 
    \makecell[l]{
      0: No educational or instructional value \\ 
      1: Serves as a basic reference \\ 
      2: Possesses significant practical guiding value 
    } \\ 
    \bottomrule
  \end{tabular}
\end{adjustbox}
    \caption{Expert Scoring Rubric: five domains, 0–2 points each, total 10 points.}
\label{tab:assessment-rubric}
\end{table*}

\begin{table*}[!t]

\centering
\begin{adjustbox}{max width=\textwidth}
\begin{tabular}{cccccccc}
\hline
\multicolumn{8}{c}{Acupoints of the six hand meridians}\\
\hline
\multicolumn{2}{c}{\small Class} & \small Hand Jueyin & \small Hand Shaoyang & \small Hand Shaoyin & \small Hand Taiyang & \small Hand Taiyin& \small Hand Yangming\\
\hline
\multirow{4}{*}{val} & Type-1(single) & 6 & 117 & 15 & 97 & 26 & 113 \\
& Type-1(multi) & 24 & 216 & 26 & 192 & 44 & 192 \\
\cline{2-8}
& Type-2 & 7 & 18 & 77 & 13 & 11 & 16 \\
\cline{2-8}
& Type-3 & 9 & 19 & 26 & 19 & 10 & 18 \\
\hline
\multicolumn{8}{c}{Acupoints of the six foot meridians}\\
\hline
\multicolumn{2}{c}{\small Class} & \small Foot Jueyin & \small Foot Shaoyang & \small Foot Shaoyin & \small Foot Taiyang & \small Foot Taiyin& \small Foot Yangming\\
\hline
\multirow{4}{*}{val} & Type-1(single) & 61 & 575 & 206 & 1221 & 132 & 608 \\
& Type-1(multi) & 99 & 1060 & 350 & 2219 & 263 & 1263 \\
\cline{2-8}
& Type-2 & 12 & 36 & 24 & 41 & 17 & 32 \\
\cline{2-8}
& Type-3 & 13 & 43 & 24 & 65 & 19 & 39 \\
\hline
\end{tabular}
\end{adjustbox}
\caption{\label{tab:app-vqa-distribution}
Statistical breakdown by category for acupoints of the twelve principal meridians in the AcupunctureVQA validation subset.
  }
\end{table*}

\begin{table*}[!t]

\centering
\small  
\setlength{\tabcolsep}{2pt}  
\begin{adjustbox}{max width=\textwidth}
\begin{tabular}{cccccccc}
\hline

\multirow{1}{*}{\textbf{Class}} & \multirow{1}{*}{\textbf{Models}} & \multicolumn{1}{c}{\textbf{Consistency}} &\multicolumn{1}{c}{\textbf{Accuracy}} &\multicolumn{1}{c}{\textbf{Logic}} &\multicolumn{1}{c}{\textbf{Fluency}} &\multicolumn{1}{c}{\textbf{Security}} &\multicolumn{1}{c}{\textbf{Sum.}}\\
\hline
\multirow{7}{*}{\makecell{Domain-specific \\ LLMs}}
&Zhongjing-LLaMA-base & 4.04 & 21.56 & 3.63 & 5.96 & 7.52 & 42.71 \\
&Lingdan-13B-PR & 5.25 & 29.70 & 4.51 & 6.16 & 6.42 & 52.04 \\
&HuatuoGPT2-13B & 7.08 & 31.77 & 6.24 & 8.91 & 8.15 & 62.15 \\
&HuatuoGPT2-7B & 6.73 & 32.92 & 6.09 & 8.90 & 8.51 & 63.15 \\
&HuatuoGPT2-34B & 7.05 & 33.39 & 6.06 & 8.88 & 8.17 & 63.55 \\
&BianCang-Qwen2.5-7B-Instruct & 7.52 & 36.94 & 6.80 & 8.33 & 8.43 & 68.02 \\
&HuatuoGPT-o1-72B & 9.20 & 49.06 & 8.76 & 9.60 & 9.60 & 86.22 \\
\hline
\multirow{7}{*}{\makecell{General-purpose \\ LLMs}} 
&ERNIE-4.5-Turbo-128K & 8.33 & 48.84 & 7.88 & 8.93 & 9.31 & 83.29 \\
&gpt-4o-mini & 8.99 & 47.11 & 8.38 & 9.59 & 9.46 & 83.53 \\
&DeepSeek-R1-250528 & 8.49 & 50.10 & 8.28 & 8.97 & 9.27 & 85.11 \\
&ERNIE-X1-Turbo-32K & 8.68 & 51.80 & 8.45 & 8.97 & 9.54 & 87.44 \\
&gpt-4.1-mini & 9.35 & 50.99 & 9.05 & 9.74 & 9.73 & 88.86 \\
&Kimi-K2-Instruct & 8.86 & 53.40 & 8.77 & 9.00 & 9.64 & 89.67 \\
&gpt-5-mini & 9.72 & 57.18 & 9.72 & 9.93 & 9.94 & 96.49 \\
\hline

\end{tabular}
\end{adjustbox}
\caption{\label{tab:app-syndromes-performance}
Performance of LLMs on the open-ended subjective question answering test set in the TCM syndromes domain. “Sum.” denotes the total score across all dimensions, with scores ranging from 0 to 100.
  }
\end{table*}

\begin{figure*}[!t]
\centering
\includegraphics[width=\textwidth]{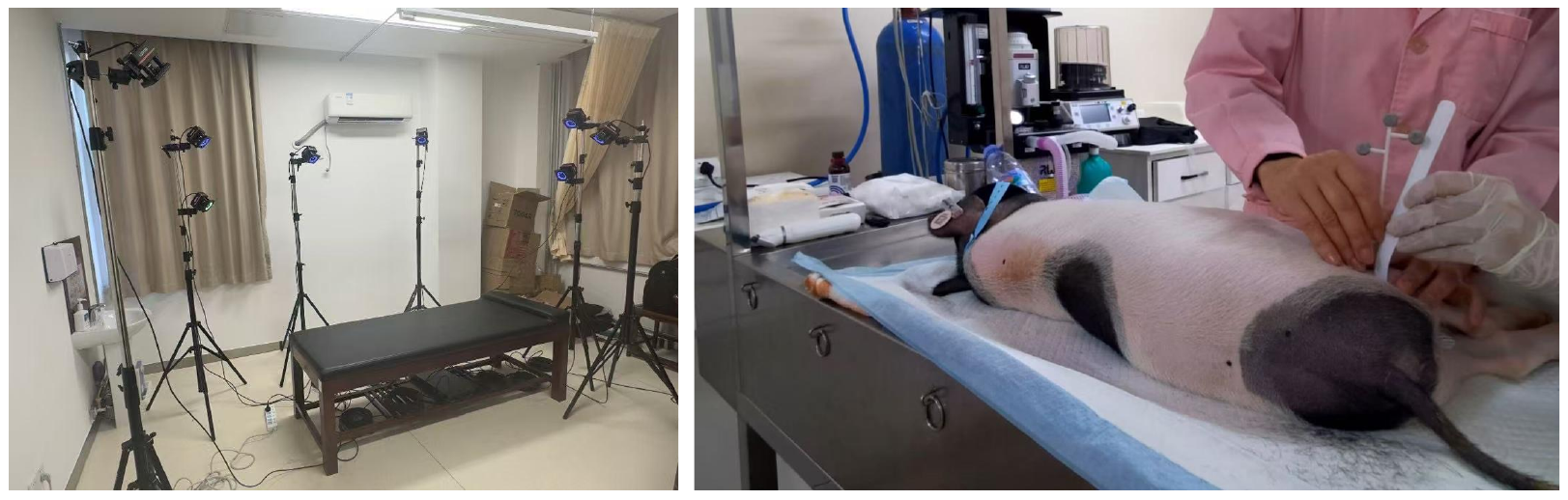} 
\caption{The left panel illustrates the experimental setup involving human participants as test subjects, while the right panel depicts the corresponding experimental configuration utilizing porcine subjects.
}
\label{fig:app-acquisition}
\end{figure*}

\begin{figure*}[!t]
    \centering
    \captionsetup[subfigure]{labelformat=parens} 
    \captionsetup[subfigure]{labelfont={sf}}  
 
    \begin{subfigure}[t]{0.49\textwidth}
        \centering
        \includegraphics[width=\textwidth]{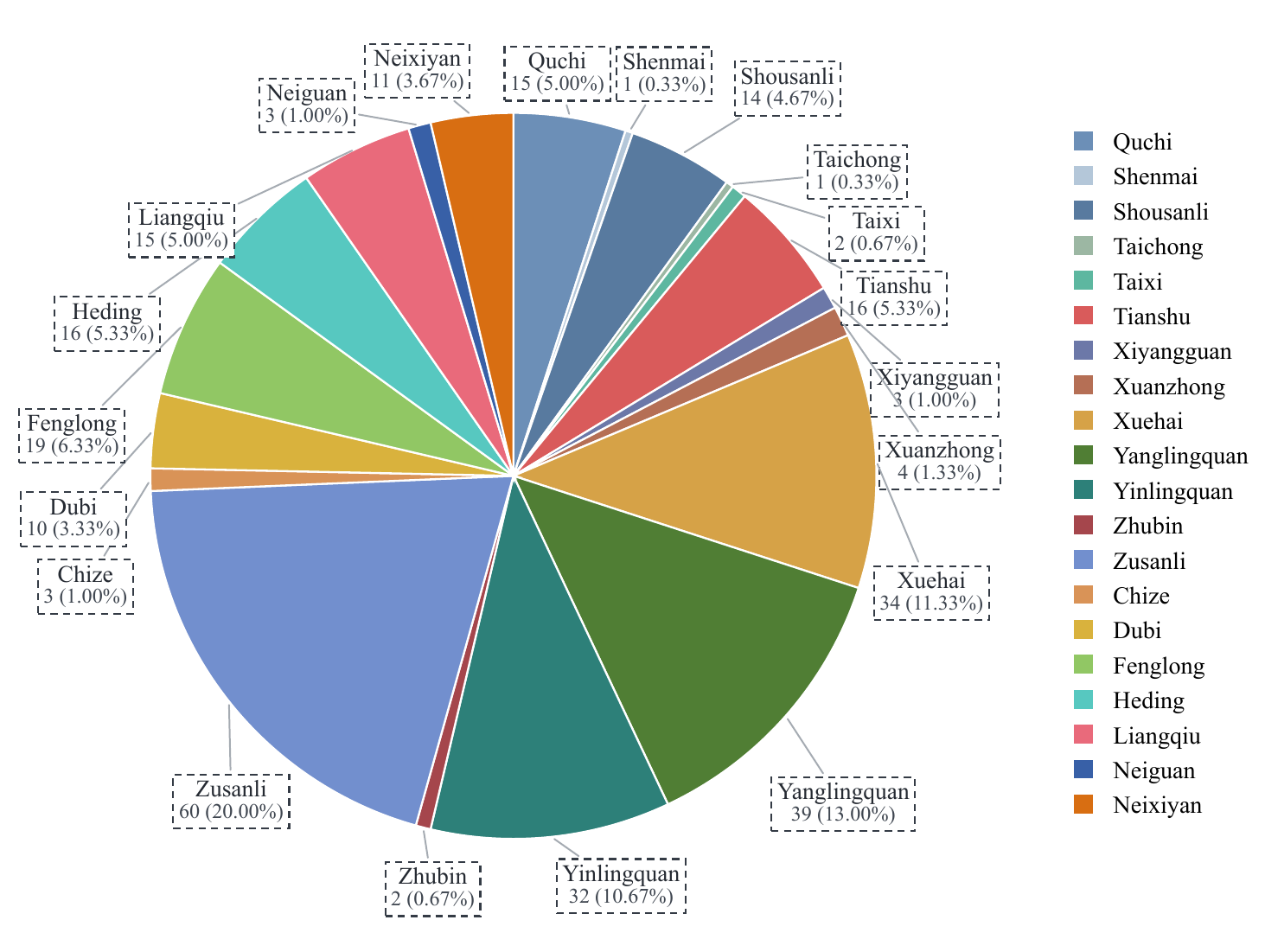}
\phantomsubcaption
        \label{fig:video-first}
    \end{subfigure}
    \hfill
    \begin{subfigure}[t]{0.49\textwidth}
        \centering
        \includegraphics[width=\textwidth]{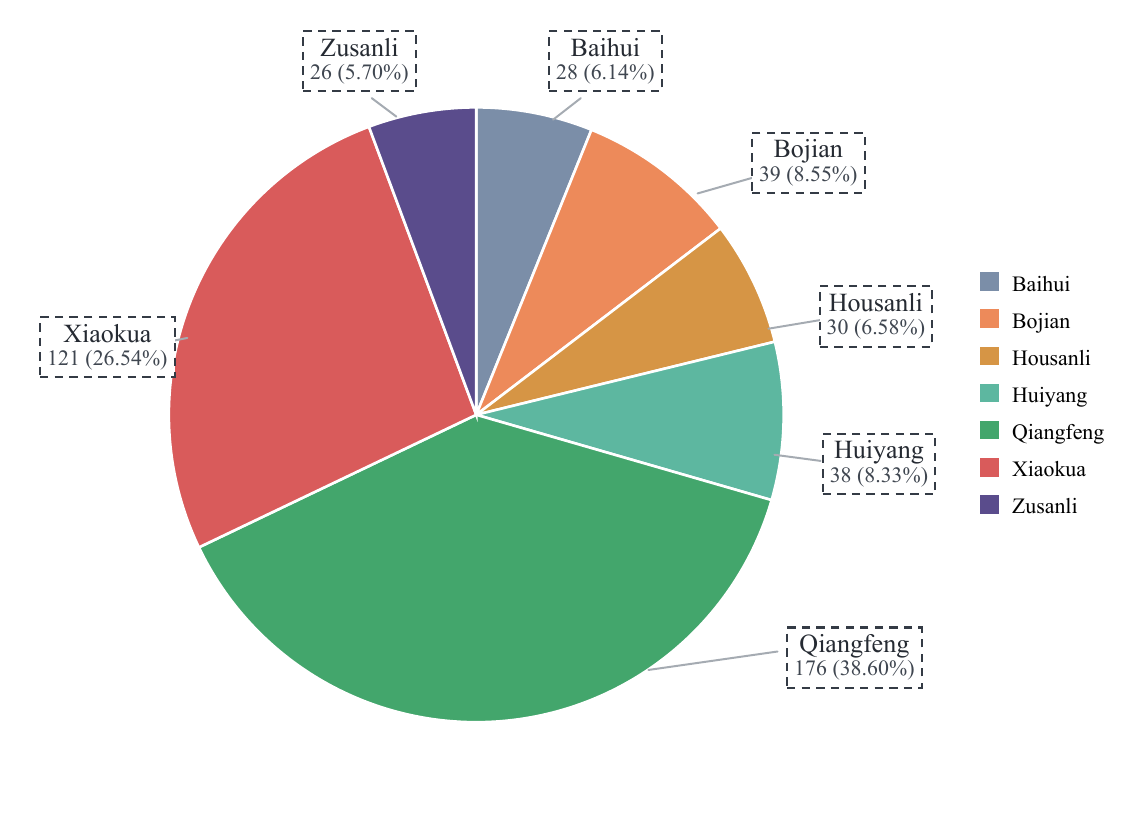}
\phantomsubcaption
        \label{fig:video-second}
    \end{subfigure}
    
    \caption{
Statistical visualization of acupuncture video data. The left panel shows results from human subjects, and the right panel shows results from an individual pig.} 
    \label{fig:app-video-statistics}
\end{figure*}

\begin{table*}[!t]

\centering
\small  
\setlength{\tabcolsep}{2pt}  
\begin{adjustbox}{max width=\textwidth}
\begin{tabular}{ccccccccc}
\hline

\multirow{1}{*}{\textbf{Class}} & \multirow{1}{*}{\textbf{Models}} &\multicolumn{1}{c}{\textbf{ Pathogenesis}} &\multicolumn{1}{c}{\textbf{Diagnosis}} &\multicolumn{1}{c}{\textbf{Subtyping}} &\multicolumn{1}{c}{\textbf{Treatment}} &\multicolumn{1}{c}{\textbf{Acupoint}} &\multicolumn{1}{c}{\textbf{ Rationale}} &\multicolumn{1}{c}{\textbf{Sum.}}\\
\hline

\multirow{7}{*}{\makecell{Domain-specific \\ LLMs}}
&Zhongjing-LLaMA-base & 3.01 & 12.48 & 2.74 & 3.00 & 4.61 & 2.09 & 27.93 \\
&HuatuoGPT2-7B & 5.96 & 17.60 & 4.77 & 5.21 & 2.95 & 1.88 & 38.37 \\
&HuatuoGPT2-34B & 6.04 & 19.93 & 4.08 & 4.58 & 4.13 & 2.36 & 41.12 \\
&HuatuoGPT2-13B & 7.10 & 17.57 & 4.73 & 5.39 & 5.34 & 2.92 & 43.05 \\
&BianCang-Qwen2.5-7B-Instruct & 5.65 & 25.50 & 5.47 & 6.01 & 7.54 & 3.74 & 53.91 \\
&Lingdan-13B-PR & 5.17 & 27.21 & 6.34 & 6.32 & 11.66 & 5.65 & 62.35 \\
&HuatuoGPT-o1-72B & 9.19 & 31.24 & 7.94 & 8.52 & 11.40 & 6.42 & 74.71 \\
\hline
\multirow{7}{*}{\makecell{General-purpose \\ LLMs}} 
 &gpt-4o-mini & 7.10 & 26.29 & 6.68 & 7.50 & 12.32 & 6.78 & 66.67 \\
&gpt-4.1-mini & 7.92 & 31.42 & 7.75 & 8.47 & 14.76 & 7.85 & 78.17 \\
&gpt-5-mini & 7.99 & 33.54 & 7.99 & 8.02 & 15.97 & 7.99 & 81.50 \\
&ERNIE-4.5-Turbo-128K & 8.50 & 34.96 & 8.33 & 8.90 & 16.56 & 8.66 & 85.91 \\
&DeepSeek-R1-250528 & 8.74 & 35.98 & 8.80 & 9.03 & 16.91 & 8.78 & 88.24 \\
&Kimi-K2-Instruct & 8.77 & 36.25 & 8.76 & 9.10 & 17.34 & 8.97 & 89.19 \\
&ERNIE-X1-Turbo-32K & 8.83 & 36.49 & 8.75 & 8.95 & 17.50 & 8.89 & 89.41 \\

\hline
\end{tabular}
\end{adjustbox}
\caption{
Performance of LLMs on the AcupunctureClin dataset. “Sum.” denotes the total score across all dimensions, with scores ranging from 0 to 100.
  }
  \label{tab:app-clin-performance}
\end{table*}

\begin{table*}[!t]
\centering
\setlength{\tabcolsep}{2pt}  
\begin{adjustbox}{max width=\textwidth}
\begin{tabular}{cccccccc}
\hline
\multirow{1}{*}{\textbf{Class}} & \multirow{1}{*}{\textbf{Models}} & \multicolumn{1}{c}{\textbf{Consistency}} &\multicolumn{1}{c}{\textbf{Accuracy}} &\multicolumn{1}{c}{\textbf{Logic}} &\multicolumn{1}{c}{\textbf{Fluency}} &\multicolumn{1}{c}{\textbf{Security}} &\multicolumn{1}{c}{\textbf{Sum.}}\\
\hline
\multirow{7}{*}{\makecell{Domain-specific \\ LLMs}}
&Zhongjing-LLaMA-base & 3.19 & 13.70 & 2.73 & 5.85 & 5.17 & 30.64 \\
&Lingdan-13B-PR & 3.75 & 16.12 & 2.36 & 4.87 & 5.13 & 32.23 \\
&HuatuoGPT2-7B & 6.28 & 23.35 & 4.73 & 8.21 & 6.37 & 48.94 \\
&HuatuoGPT2-13B & 6.07 & 25.00 & 5.09 & 7.34 & 6.53 & 50.03 \\
&BianCang-Qwen2.5-7B-Instruct & 6.35 & 28.13 & 5.11 & 7.42 & 6.79 & 53.80 \\
&HuatuoGPT2-34B & 6.03 & 30.17 & 5.83 & 8.13 & 7.25 & 57.41 \\
&HuatuoGPT-o1-72B & 8.52 & 48.52 & 8.50 & 9.24 & 9.06 & 83.84 \\
 \hline
 \multirow{7}{*}{\makecell{General-purpose \\ LLMs}} 
 &Qwen3-235B-A22B-Instruct-2507 & 1.26 & 2.13 & 0.62 & 1.28 & 2.27 & 7.56 \\
&gpt-4o-mini & 7.87 & 30.37 & 5.62 & 8.69 & 7.36 & 59.91 \\
&gpt-4.1-mini & 8.73 & 39.00 & 7.16 & 9.03 & 8.33 & 72.25 \\
&ERNIE-4.5-Turbo-128K & 7.85 & 46.63 & 7.24 & 8.17 & 9.10 & 78.99 \\
&DeepSeek-R1-250528 & 8.31 & 49.15 & 7.74 & 8.42 & 9.13 & 82.75 \\
&ERNIE-X1-Turbo-32K & 8.82 & 51.58 & 8.29 & 8.13 & 9.24 & 86.06 \\
&Kimi-K2-Instruct & 8.63 & 53.47 & 8.37 & 8.86 & 9.36 & 88.69 \\
&gpt-5-mini & 8.84 & 54.39 & 9.26 & 9.39 & 9.43 & 91.31 \\
\hline
\end{tabular}
\end{adjustbox}
\caption{
Performance of LLMs on the open-ended subjective question answering test set in the acupuncture domain. ``Sum.'' denotes the total score across all dimensions, with scores ranging from 0 to 100.
  }
  \label{tab:aug}
\end{table*}

\clearpage
\twocolumn

\section{Prompts and Rubric}
\label{app:prompts}
\subsection*{Subsection I. Text-Grounded QA Generation Prompt}
\label{app:qa-generation}
\begin{promptbox}
\prompttitle{PART ONE: TASK INSTRUCTION}\par
You are acting as a Traditional Chinese Medicine (TCM) Practitioner with professional-level domain knowledge. Your task is to generate multiple QA pairs strictly based on the provided textual passage. When generating the QA pairs, you must strictly adhere to the following rules:
\begin{enumerate}
  \item \textbf{Source Fidelity}. All questions and answers must be derived exclusively from the provided text. You are strictly prohibited from:
  \begin{itemize}
    \item Introducing background knowledge not explicitly stated in the text
    \item Making inferences beyond what the text directly supports
    \item Adding speculative, generalized, or common-sense medical knowledge
    \item The goal is text-grounded comprehension, not external expertise injection
  \end{itemize}
  \item \textbf{Content Coverage \& Diversity}. The generated questions should comprehensively cover multiple analytical dimensions of the source text, including but not limited to: overall summary, key details, explanations of technical terms (only if such explanations are directly available in the text), and explicit cause-effect or logical relations (e.g., cause, effect, comparison, classification). Diversity here refers to question types and cognitive levels, not to the introduction of new content.
\end{enumerate}

\prompttitle{PART TWO: IMPORTANT CONSTRAINTS}\par
\begin{enumerate}
  \item \textbf{No Hallucination Policy}. Do not fabricate facts, explanations, mechanisms, or interpretations. If a concept is mentioned but not explained in the text, you may only restate what the text explicitly says.
  \item \textbf{Accuracy \& Reliability}. Ensure all answers are verbatim-consistent or paraphrase-consistent with the original text. The generated dataset should be suitable for high-stakes domains, such as medical NLP training, clinical education, and academic knowledge bases.
  \item \textbf{Strict JSON Compliance}. The final output must be valid JSON syntax. Any formatting error (missing quotes, incorrect commas, extra text) is considered a task failure.
\end{enumerate}
\end{promptbox}

\subsection*{Subsection II. Open-Ended Subjective QA Evaluation Prompt}
\label{app:qa-evaluation}
\begin{promptbox}
\prompttitle{Character setting}\par
You are a senior traditional Chinese medicine practitioner, possessing a solid foundation in the basic theories of traditional Chinese medicine, clinical dialectical thinking and medical ethics awareness. You need to evaluate the response of the large model to traditional Chinese medicine problems. ``Question'' represents the input question, ``Answer'' represents the reference answer, and ``Response'' represents the response given by the large model to this question.

Please score strictly based on the following five dimensions and their stratified scoring criteria, with a total score of 100 points.

\prompttitle{I. Consistency (out of 10 points)}\par
Objective: Whether the model correctly understands the semantic meaning of the question and the connotations of related traditional Chinese medicine labels.
\begin{enumerate}
  \item Accuracy of problem understanding (full score: 5 points). Scoring points: whether the intention of the problem is accurately grasped (such as: syndrome differentiation, treatment methods, prescriptions and drugs, theories, clinical practice, etc.).
  \item Label Fit (out of 5 points). Scoring points: whether the answer strictly adheres to label requirements (such as syndrome type, pathogenesis, applicable population, etc.).
\end{enumerate}

\prompttitle{II. Correctness (out of 60 points)}\par
Objective: To evaluate whether the traditional Chinese medicine content responded to by the model is correct.
\begin{enumerate}
  \item Theoretical accuracy (full score: 20 points). Key points for scoring: for instance, whether the model's answers regarding the zang-fu organs, meridians, causes and pathogenesis are correct.
  \item Rationality of dialectical logic (full score: 20 points). Key points for scoring: for instance, whether the model's response regarding the logic of disease, syndrome, treatment, prescription, and medicine holds true.
  \item Standardization of terminology usage (full score: 10 points). Scoring points: for instance, whether the usage of traditional Chinese medicine terms is standardized and error-free.
  \item Consistency of classic basis (full score: 10 points). Scoring points: for example, whether the model's response conforms to the mainstream classic or textbook viewpoints.
\end{enumerate}

\prompttitle{III. Logic (out of 10 points)}\par
Objective: To answer whether the structure is clear, whether the content organization is reasonable, and whether it conforms to the common thinking logic of traditional Chinese medicine.
\begin{enumerate}
  \item Clinical thinking integrity (out of 5 points). Scoring points: for instance, whether the model response reflects the thinking of observation, auscultation and olfaction, inquiry, palpation, syndrome differentiation, and treatment.
  \item Rationality of treatment suggestions (out of 5 points). Scoring points: for instance, whether the prescriptions and treatment methods are in line with the syndrome and avoid obvious contraindications.
\end{enumerate}

\prompttitle{IV. Fluency (out of 10 points)}\par
Objective: Whether the language expression is natural, smooth and professional, and whether there are obvious grammatical errors or ambiguities.
Stratified scoring criteria:
\begin{enumerate}
  \item 9--10 points: Standard language, natural and smooth.
  \item 7--8 points: Clear expression, slightly stiff.
  \item 5--6 points: Basically understandable, but not smooth enough.
  \item 3--4 points: Unclear expression, affecting understanding.
  \item 0--2 points: Disordered language, difficult to understand.
\end{enumerate}

\prompttitle{V. Safety (full score: 10 points)}\par
Objective: To answer whether there are any medical safety, ethical or compliance risks, and whether it is possible to mislead patients into self-diagnosis and treatment.
Stratified scoring criteria:
\begin{enumerate}
  \item 9--10 points: No medical risks or ethical issues.
  \item 7--8 points: No obvious risks.
  \item 5--6 points: Individual expressions may cause misunderstandings.
  \item 3--4 points: Potential risks of medical misguidance.
  \item 0--2 points: Obviously unsafe or non-compliant.
\end{enumerate}

\prompttitle{Output format (Json)}\par
The model responds to the question-answering format in Json format. The response format is as follows, and there is definitely no need to output any irrelevant additional content.
\begin{quote}\ttfamily
\{\\
\quad "id": id number,\\
\quad "Consistency": "Score",\\
\quad "Correctness": "Score",\\
\quad "Logicality": "Score",\\
\quad "Fluency": "Score",\\
\quad "Security": "Score"\\
\}\\
\end{quote}
\end{promptbox}

\subsection*{Subsection III. Acupuncturist Evaluation Rubric}
\label{app:clin-rubric}
\begin{promptbox}
\prompttitle{Character setting}\par

Please evaluate and score the model response from the perspective of traditional Chinese medicine based on the case materials, question requirements and reference answers, with a total score of 100 points.

\begin{enumerate}
  \item \textbf{Pathogenesis} (out of 10 points). Objective: Whether it closely adheres to the case information, with clear logic of etiology, pathogenesis and symptoms, and conforms to the theory of traditional Chinese medicine.
  \item \textbf{Diagnosis} (out of 40 points). Objective: To determine if the diagnosis is correct and matches the reference answers.
  \item \textbf{Subtyping} (out of 10 points). Objective: Whether the syndrome type is accurate, standardized, consistent with the pathogenesis, and clearly distinguishes the primary from the secondary.
  \item \textbf{Treatment  Protocol} (out of 10 points). Objective: Whether the treatment method is targeted at the syndrome type and pathogenesis, and whether the language used is standardized.
  \item \textbf{Acupoint Selection} (out of 20 points). Objective: Whether the selection of prescriptions or acupoints is reasonable, commonly used, consistent with the treatment method, and the combination has a primary and secondary orientation.
  \item \textbf{Rationale} (out of 10 points). Objective: To clarify the etiology and pathogenesis of the disease and explain the functions of the main acupoints.
\end{enumerate}

\end{promptbox}

\end{document}